\documentclass[12pt]{article}
\usepackage{putex}
\usepackage{graphicx}
\usepackage{epstopdf}

\allowdisplaybreaks[1]

\begin{document}

\preprint{hep-th/0609137 \\ PUPT-2207}

\institution{PU}{Joseph Henry Laboratories, Princeton University, Princeton, NJ 08544}

\title{Stability of strings binding heavy-quark mesons}

\authors{Joshua J. Friess, Steven S. Gubser, Georgios Michalogiorgakis, and \\[10pt] Silviu S. Pufu}

\abstract{\noindent We investigate the stability against small deformations of strings dangling into $AdS_5$-Schwarzschild from a moving heavy quark-anti-quark pair.  We speculate that emission of massive string states may be an important part of the evolution of certain unstable configurations.}


\maketitle
\tableofcontents

\providecommand{\abs}[1]{\lvert#1\rvert}
\providecommand{\norm}[1]{\lVert#1\rVert}
\def\im{\operatorname{Im}}
\def\re{\operatorname{Re}}

\section{Introduction}

In \cite{Herzog:2006gh,Peeters:2006iu,Liu:2006nn,Chernicoff:2006hi}, string configurations in $AdS_5$-Schwarzschild were exhibited which were argued to describe the propagation of a heavy-quark meson through a thermal quark-gluon-plasma.  Briefly, the conclusion of these works is that quarkonium systems can propagate without experiencing any drag force, provided they are small enough and their velocity is not too high.  Part of the interest of this topic stems from measurements at RHIC of $J/\psi$ suppression relative to binary scaling expectations \cite{PereiraDaCosta:2005xz}, which is not as severe as expected based on models assuming color screening (see for example \cite{Muller:2006ee}).  Could the no-drag property predicted using AdS/CFT have to do with the weaker than expected $J/\psi$ suppression?

The present paper aims to address more modest questions:
 \begin{enumerate}
  \item Are the string configurations studied in \cite{Peeters:2006iu,Liu:2006nn,Chernicoff:2006hi} stable against linear perturbations?\label{LocallyStable}
  \item When / if one of these configurations is unstable, what does it evolve into?
 \end{enumerate}
Along the way we will encounter an analytic expression for the shape of the string describing a moving quarkonium system.

A related work \cite{Argyres} appeared recently which emphasizes global comparisons of different branches of the configuration space of strings attached to the same flavor brane.  Our analysis, along the lines of point~\ref{LocallyStable}, is complementary in that stability is examined solely from the point of view of small deformations about a given classical solution.

The rest of the paper is organized as follows.  In section~\ref{PERTURB} we give an account of linear perturbations.  In section~\ref{EVOLVE} we speculate about the evolution of unstable perturbations.  We end in section~\ref{CONCLUSION} with a brief summary of our conclusions.

\section{Perturbations of a quarkonium string configuration}
\label{PERTURB}

The metric of $AdS_5$-Schwarzschild is
\eqn{Metric}{
 ds^2 = {L^2 \over
z^2} \left[ - \left( 1 - {z^4 \over z_H^4} \right) dt^2 +
\left(dx^1\right)^2 + \left(dx^2\right)^2 + \left(dx^3\right)^2 + {1
\over 1 - {z^4 \over z_H^4}} dz^2 \right] \,.
}
 We will assume that
while the ends of the string are separated in the $x^2$~direction by
coordinate distance $\ell$, they are constrained to move with
velocity $v$ in the $x^1$~direction.  Thus we are already restricting ourselves to a special case of the analysis of \cite{Liu:2006nn,Chernicoff:2006hi}, which is roughly the configuration of \cite{Peeters:2006iu} without spin.  It was found in \cite{Liu:2006nn,Chernicoff:2006hi} that no
solutions are possible unless the separation between the two ends is
less than some critical value $\ell<\ell_{max}$.  If the separation
is smaller than $\ell_{max}$, then two solutions are possible.  We will argue that only one of them is stable against small perturbations, namely the one that dangles less far into $AdS_5$-Schwarzschild and has lower energy.

We work in the string's equilibrium rest frame: that is, we boost $x^1 \to \gamma (x^1 + v t)$ and $t \to
\gamma (t + x^1 v)$ to get
\eqn{BoostedMetric}{
 ds^2 =
G_{\mu\nu} dx^\mu dx^\nu = {L^2 \over z^2} \bigg[ - \left( 1 - {z^4
\over z_H^4} \gamma^2 \right) dt^2 + 2 {z^4 \over z_H^4} \gamma^2 v
dt dx^1 +\cr \left(1 + {z^4 \over z_H^4} \gamma^2 v^2 \right)
\left(dx^1\right)^2 + \left(dx^2\right)^2 + \left(dx^3\right)^2 + {1
\over 1 - {z^4 \over z_H^4}} dz^2 \bigg] \,.
}
The string's classical dynamics is described by the Nambu-Goto action:
\begin{align}
 S = - {1 \over 2 \pi \alpha'} \int d^2 \sigma
\sqrt{- \det g_{\alpha\beta}} && g_{\alpha\beta} = G_{\mu\nu}
\partial_\alpha X^\mu \partial_\beta X^\nu\label{NambuGoto}\,,
\end{align}
  where $X^\mu$ represents the embedding of the string into
$AdS_5$-Schwarzschild and
$\sigma^\alpha = (\tau,\sigma)$ are the worldsheet coordinates. The
position of the string can then be described by
\begin{equation} \label{StringCoord}
  X^\mu = \begin{pmatrix}
    X^0(\tau, \sigma) & X^1(\tau, \sigma)& X^2 (\tau, \sigma)& X^3(\tau, \sigma)& Z(\tau, \sigma)
  \end{pmatrix} \,.
\end{equation}
We choose ``static'' gauge, $X^0(\tau, \sigma) = \tau$.  An obvious
completion of this gauge choice is $X^2(\tau,\sigma) = \sigma$, but
it turns out that $Z(\tau,\sigma) = \sigma$ is a better one because
then we are able to express the equilibrium configuration of the
string in closed form (see \eno{x2Equilibrium}).  An odd feature of
this gauge choice is that it covers only half the string:
$X^2(\tau,\sigma)$ is a double-valued function.  In exploring
perturbations we must therefore patch together two half-solutions at
the string midpoint.

\begin{figure}
\begin{center}
\includegraphics[width=3.1in]{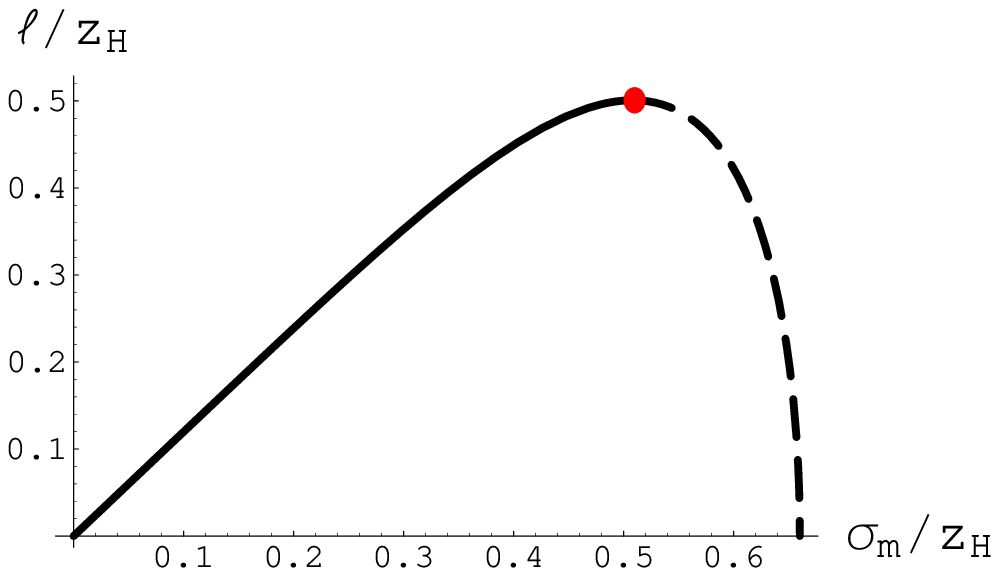}\quad\includegraphics[width=2.3in]{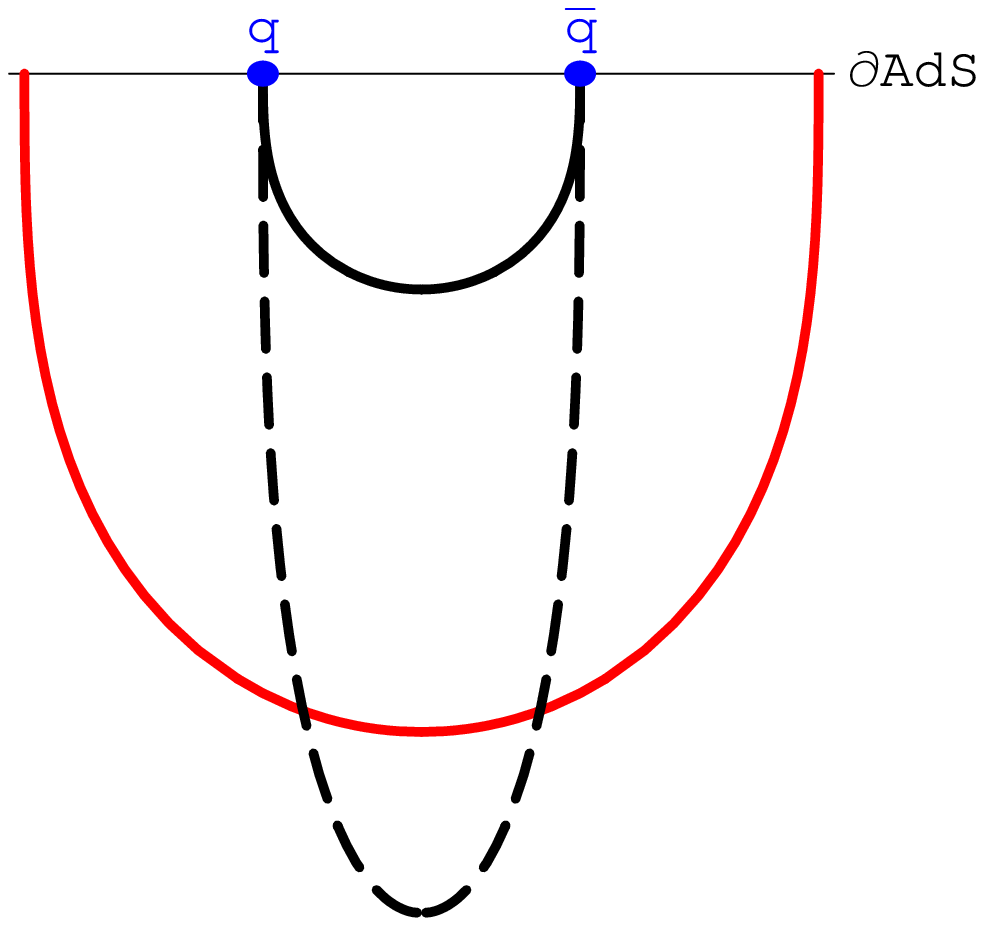}
\end{center}
 \begin{picture}(0,0)(0,0)
  \put(120,0){\Large (A)}
  \put(335,0){\Large (B)}
 \end{picture}
\caption{A) The separation $\ell$ as a function of $\sigma_m$ for $ v =
0.9$.  (See \eno{separation}.)  As explained below
\eno{x2Equilibrium}, $\sigma_m$ is the maximum value of $z$ along
the string.  The solid part of the plot corresponds to stable
configurations; the dashed part corresponds to unstable
configurations; and the red dot corresponds to $\sigma = \sigma_{\rm
max}$, where the separation between the quark and anti-quark is
maximized.  B) A cartoon of string configurations corresponding to three points on the curve from A.  For any separation of quark and anti-quark less than the maximum (corresponding to the red curve), there is a stable string configuration (solid) and an unstable one (dashed).} \label{fig:lvssigmam}
\end{figure}

With the gauge choice explained in the previous paragraph, the Nambu-Goto lagrangian density is
\eqn{LagrDensity}{
  \mathcal{L} = -\sqrt{ { \left(1-v^2\right) \left(z_H^4-\sigma^4\right)^2 \left({d X^1 \over d \sigma} \right)^2 + \left[\left(1-v^2\right) z_H^4 - \sigma^4 \right] \left[z_H^4 + \left(z_H^4 - \sigma^4 \right) \left(\left({d X^2\over d \sigma}\right)^2 + \left({d X^3\over d \sigma} \right)^2 \right) \right] \over \left(1-v^2\right) z_H^4 \sigma^4 \left( z_H^4 - \sigma^4 \right) }
  }\,.
}
 Under the assumption of an everywhere smooth string configuration, one may demonstrate that $X^1 = X^3 \equiv 0$ \cite{Chernicoff:2006hi}.  If we solve the $Z$-equation of
motion we get, for the left half of the string (where $X^2<0$),
\eqn{Equilibrium}{
  X^\mu = \begin{pmatrix}
    \tau & 0 & X_e^2 (\sigma)& 0& \sigma
  \end{pmatrix} \,,
}
 with
\begin{multline} \label{x2Equilibrium}
X_e^2(\sigma) = {\sigma^3 \sqrt{(1-v^2) z_H^4 - \sigma_m^4} \over 3
\sqrt{1-v^2} z_H^2 \sigma_m^2} F_1\left({3 \over 4}; {1 \over 2}, {1
\over 2}; {7 \over 4}; {\sigma^4 \over z_H^4}, {\sigma^4 \over
\sigma_m^4} \right) -\\* {\sigma_m \sqrt{(1-v^2) z_H^4 - \sigma_m^4}
\over 3 \sqrt{1-v^2} z_H^2} \sqrt{\pi} \Gamma\left({7 \over
4}\right) {}_2F_1 \left( {3 \over 4}, {1 \over 2}; {5 \over 4};
{\sigma_m^4 \over z_H^4} \right) \,,
\end{multline}
where  $F_1$ is the Appell hypergeometric function.  The right half
of the string is described by the same solution \eno{x2Equilibrium}
with an overall sign flip, $X^2_e \to -X^2_e$.  Here,
$X_e^2(\sigma_m) = 0$, so $\sigma_m$ can be understood as the
largest $z$-coordinate achieved by the string.
As noted in
\cite{Chernicoff:2006hi, Liu:2006nn}, the above solution only exists
as long as the string doesn't get too close to the horizon, namely
as long as
\begin{equation} \label{zv}
\sigma_m \leq z_v \equiv z_H (1-v^2)^{{1 \over 4}} \,.
\end{equation}
We have thus parameterized our solution by $\sigma_m$, with $0 \leq
\sigma_m \leq z_v$.  We can relate $\sigma_m$ to the coordinate
distance between the string's endpoints by writing $\ell = - 2
X_e^2(0)$ which gives
\begin{equation} \label{separation}
  \ell = {2 \sqrt{\pi} \sigma_m \sqrt{(1-v^2) z_H^4 - \sigma_m^4} \Gamma\left({7 \over 4} \right) \over 3
  \sqrt{1-v^2} z_H^2 \Gamma\left({5 \over 4} \right)} {}_2F_1 \left( {3 \over 4}, {1 \over 2}; {5 \over 4};
{\sigma_m^4 \over z_H^4} \right)\,.
\end{equation}
An equivalent result to \eno{separation} has already appeared in \cite{Avramis:2006em}.  A plot of $\ell$ as a function of $\sigma_m$ can be seen in
figure~\ref{fig:lvssigmam}, where $v = 0.9$.
For this particular value of $v$ the maximum $\ell_{max}$ is
attained at $\sigma_m = \sigma_{\rm max} \approx 0.51 z_H$.  As we shall
see, all configurations that have $\sigma_m<\sigma_{\rm max}$ are stable
with respect to small perturbations, and all configurations with
$\sigma_m>\sigma_{\rm max}$ are unstable.

To prove this claim, we do a linearized stability analysis of the
equilibrium configurations given by \eqref{Equilibrium}.  We write
\begin{equation} \label{perturbed}
  X^\mu = \begin{pmatrix}
    \tau& \delta X^1 (\tau, \sigma) & X_e^2 (\sigma) + \delta X^2 (\tau, \sigma) & X^3 (\tau, \sigma) & \sigma + \delta Z(\tau, \sigma)
  \end{pmatrix} \,,
\end{equation}
where we have set $\delta X^2 = 0$ because any
non-vanishing $\delta X^2$ corresponds to reparameterizations of the
in-plane perturbations.  With this choice, and denoting $\delta X^1
\equiv \delta X_\parallel$ and $\delta X^3 \equiv \delta X_\perp$,
we derive the linearized equations of motion that follow from the
action \eqref{NambuGoto}:
\begin{gather}
\left[ -\begin{pmatrix}   m_\parallel & 0 \\ 0 & m_Z
\end{pmatrix} \partial_\tau^2  + \partial_\sigma \begin{pmatrix}   g_\parallel & 0 \\ 0 & g_Z
\end{pmatrix} \partial_\sigma + \begin{pmatrix}   0 & B \partial_\tau  \\ -B \partial_\tau& f_Z
\end{pmatrix} \right] \begin{pmatrix}
    \delta X_\parallel \\  \delta Z
\end{pmatrix} = 0 \label{linearizedeqsx1z}\\*[2\jot]
\left( - m_\perp \partial_\tau^2  + \partial_\sigma g_\perp
\partial_\sigma \right) \delta X_\perp = 0 \label{linearizedeqx3}
\end{gather}
where
\begin{align}
 m_\parallel &= -{\sqrt{1- {\sigma^4 \over z_H^4}} \over \sigma^2 \left( 1- {\sigma^4 \over z_v^4} \right)
 \sqrt{1- {\sigma^4 \over \sigma_m^4}}} \label{mparDef}\\
 m_\perp &= -{1 \over \sigma^2 \sqrt{1-{\sigma^4 \over z_H^4}} \sqrt{1- {\sigma^4 \over
 \sigma_m^4}}} \label{mperpDef}\\
 m_Z &= -{\sigma^2 \left(1- {\sigma_m^4 \over z_v^4} \right) \over \sigma_m^4 \left( 1- {\sigma^4 \over z_v^4}\right) \left( 1- {\sigma^4 \over z_H^4}\right)^{3/2} \sqrt{1- {\sigma^4 \over
 \sigma_m^4}}} \label{mZDef}\\
 g_\parallel &= -{\left( 1- {\sigma^4 \over z_H^4}\right)^{3/2} \sqrt{1- {\sigma^4 \over
 \sigma_m^4}} \over \sigma^2 \left(1 - {\sigma^4 \over z_v^4}
 \right)}\label{gparDef}\\
 g_\perp &=-{\sqrt{1- {\sigma^4 \over z_H^4}} \sqrt{1- {\sigma^4 \over
 \sigma_m^4}} \over \sigma^2}\label{gperpDef}\\
 g_Z &= - {\sigma^2 \left( 1- {\sigma_m^4 \over z_v^4}\right) \sqrt{1- {\sigma^4 \over
 \sigma_m^4}} \over  \sigma_m^4 \left( 1- {\sigma^4 \over z_v^4}\right) \sqrt{1- {\sigma^4 \over
 z_H^4}}}\label{gZDef}\\
 B &= {4 v \sigma^5 \left( 1- {\sigma_m^4 \over z_v^4}\right) \over \sigma_m^4 z_H^4 (1-v^2) \left( 1- {\sigma^4 \over z_v^4}\right)^2 \sqrt{1- {\sigma^4 \over
 z_H^4}} \sqrt{1- {\sigma^4 \over
 \sigma_m^4}}}\label{BDef}\\
 f_Z &= {\left( 1- {\sigma_m^4 \over z_v^4}\right) A  \over z_H^4 \left(1-v^2\right) \left( 1- {\sigma^4 \over z_v^4}\right)^2 \left( 1- {\sigma^4 \over z_H^4}\right)^{5/2} \sqrt{1- {\sigma^4 \over
 \sigma_m^4}}  } \label{fZDef}\,,
\end{align}
with
\begin{multline} \label{Adef}
  A = 14 \left(1-v^2\right){\sigma^8 \over \sigma_m^8} - 6 \left(2-v^2\right){\sigma^{12} \over z_H^4\sigma_m^8} - 2{\sigma^{16} \over z_H^8 \sigma^8} +\\* 2 \left(1-v^2\right){z_H^4 \over \sigma_m^4} -10 \left(2-v^2\right){\sigma^4 \over \sigma_m^4} + 18 {\sigma^8 \over z_H^4
  \sigma_m^4}\,.
\end{multline}

First note that equation \eqref{linearizedeqx3} for $\delta X_\perp$
decouples and simplifies to
\begin{equation} \label{x3eqsimp}
 \left[ - \partial_\tau^2 + \left( 1 - {\sigma^4
\over z_H^4} \right) \left( 1- {\sigma^4 \over \sigma_m^4} \right)
\partial_\sigma^2 -{2 \over \sigma} \left(1- {\sigma^8 \over
z_H^4 \sigma_m^4}\right)
\partial_\sigma \right] \delta X_\perp = 0
\end{equation}
Callan and G\"{u}ijosa have looked at a similar equation in
\cite{Callan:1999ki}, where they only considered the pure-AdS case.
It is easy to check that our equation \eqref{x3eqsimp} reduces to
equation (7) of \cite{Callan:1999ki}, provided we take the pure-AdS
limit $z_H \to \infty$.

As we shall see, it follows from equation \eqref{x3eqsimp} that all
transverse normal modes $\delta X_\perp(\tau, \sigma) = \re
\left\{\delta X_\perp(\sigma) e^{-i \omega \tau}\right\}$ have
$\omega^2>0$, and so the equilibrium configuration is stable with
respect to small perturbations in the $x^3$~direction. We can see
this directly by solving the above equation, but we choose a
different approach: we look at how the hamiltonian of the system
changes when we make \emph{time-independent} perturbations of the
shape of the string in the $x^3$~direction.  More explicitly, if the
hamiltonian always increases when we statically perturb the shape of
the string, then it must be that the equilibrium configuration of
the string is locally stable.  Physically, the situation is
analogous to having a particle at the minimum of a potential well.
If we slightly change its position, the particle would tend to go
back to the lowest potential energy configuration.

In our case, the hamiltonian density is
\begin{equation}\label{HamiltDensity}
\mathcal{H} = {\partial \mathcal{L} \over \partial_\tau X^i}
\partial_\tau X^i + {\partial \mathcal{L} \over \partial_\tau Z}
\partial_\tau Z - \mathcal{L}\,,
\end{equation}
and, if we assume static configurations of the form
\begin{equation} \label{static}
  X^\mu = \begin{pmatrix}
    \tau& 0 & X_e^2 (\sigma) & \delta X_\perp(\sigma)& 0
  \end{pmatrix}
\end{equation}
we get
\begin{equation} \label{HamiltXperp}
  \mathcal{H} = {\left(z_v^4 - \sigma^4\right) \sigma_m^2 \over z_H^2 \sigma^2 \left(1 - v^2\right) \sqrt{\left(\sigma_m^4 - \sigma^4\right) \left(z_H^4 - \sigma^4\right)}} + {\sqrt{\left(\sigma_m^4 - \sigma^4\right) \left(z_H^4 - \sigma^4\right)}\over 2 z_H^2 \sigma^2 \sigma_m^2} \delta X_\perp'^2 + O\left(\abs{\delta X_\perp}^3\right)\,,
\end{equation}
which is indeed positive for any small $\delta X_\perp(\sigma)$.
Hence all normal modes in the $x^3$~direction have $\omega^2>0$.
The question remains whether or not the in-plane normal modes given
by equation \eqref{linearizedeqsx1z} share the same feature.

A hamiltonian approach similar to \eno{HamiltXperp} could also be
used to explore in-plane normal modes, but we proceed instead to
solve the coupled equations in \eqref{linearizedeqsx1z} directly.
This allows us to determine the lowest normal mode explicitly.
Because we're mostly interested in the unstable regions, we look for
solutions to \eqref{linearizedeqsx1z} of the form
\begin{align}
  \delta X_\parallel(\tau, \sigma) &= \re \left\{\delta X_\parallel(\sigma) e^{\Omega \tau}
  \right\}\label{Xparnormalmodes}\\*
  \delta Z(\tau, \sigma) &= \re \left\{ \delta Z(\sigma) e^{\Omega \tau}\right\}\label{Znormalmodes}
  \,.
\end{align}
Plugging this ansatz into \eqref{linearizedeqsx1z}, we obtain the
eigenvalue problem:
\begin{equation} \label{evalue}
 \left[ \partial_\sigma \begin{pmatrix}   g_\parallel & 0 \\ 0 & g_Z
\end{pmatrix} \partial_\sigma + \begin{pmatrix}   0 & B \Omega \\ -B \Omega & f_Z
\end{pmatrix} \right] \begin{pmatrix}
    \delta X_\parallel \\  \delta Z
\end{pmatrix} = \Omega^2 \begin{pmatrix} m_\parallel & 0 \\ 0 & m_Z \end{pmatrix} \begin{pmatrix}
    \delta X_\parallel \\  \delta Z
\end{pmatrix}\,.
\end{equation}
The eigenfrequencies $\omega = i \Omega$ can be found by solving
these equations and imposing appropriate boundary conditions at
$\sigma = 0$ and $\sigma = \sigma_m$.

At $\sigma = 0$, equations \eqref{evalue} give the following
asymptotic relations for the solutions $\delta X_\parallel (\sigma)$
and $\delta Z (\sigma)$:
\begin{align}
\delta X_\parallel (\sigma) &= \alpha_1 - {1 \over 2} \alpha_1
\Omega^2 \sigma^2 + \alpha_2 \sigma^3 +
O\left(\sigma^4\right)\label{BdyAsympX}\\*
\delta Z (\sigma) &=
{\beta_1 \over \sigma^2} - {1 \over 2} \beta_1 \Omega^2 + \beta_2
\sigma + O\left(\sigma^2\right)\label{BdyAsympZ}\,,
\end{align}
where the choice of the four complex constants $\alpha_1$,
$\alpha_2$, $\beta_1$, and $\beta_2$ determines the whole series
expansion. The boundary conditions that we want to impose are
\begin{align}
  \delta X_\parallel (0) = 0 && \delta Z(0) = 0\label{BdyBCs}\,,
\end{align}
which imply setting $\alpha_1 = \beta_1 = 0$.

The first consequence of imposing these boundary conditions is that
all in-plane normal modes have real $\Omega^2$.  Indeed, if without
loss of generality we assume that $\beta_2$ is real, the reality of
$\Omega^2$ follows from examining the leading order behavior of the
imaginary part of the $\delta Z$ equation in
\eqref{linearizedeqsx1z}:  $\im\left\{\text{LHS}\right\} = - C_1
\sigma^8 \im \left\{\Omega \alpha_2 \right\} +
O\left(\sigma^9\right)$, whereas $\im\left\{\text{RHS}\right\} = C_2
\beta_2 \sigma^3 \im \left\{\Omega^2\right\} +
O\left(\sigma^4\right)$, for some non-zero real constants $C_1$ and
$C_2$. Since the leading order behavior should be the same on both
sides, we conclude that $\im \left\{\Omega^2\right\}=0$, so
$\Omega^2$ is real. Moreover, if $\Omega$ itself is real, as it is
the case for the unstable modes, we further have that $\delta
X_\parallel$ and $\delta Z$ are real \emph{everywhere}, because they
satisfy the differential equation \eqref{linearizedeqsx1z} which in
this case has real coefficients. Being interested in the unstable
modes, we will henceforth assume that all these quantities are real.

Imposing boundary conditions at $\sigma = \sigma_m$ is slightly more
complicated, because we have to patch smoothly the two halves of the
string.  In order to do this, we first express $\delta X_\parallel$
and $\delta Z$ not in terms of our usual variable $\sigma$, but
rather in terms of the equilibrium coordinate $X_e^2$, whose
expression in terms of $\sigma$ is given in \eqref{x2Equilibrium}
for the left half of the string.  The patching conditions then read
\begin{align}
  \left. \delta X_\parallel \right|_{X_e^2 = 0_{-}} &= \left. \delta X_\parallel \right|_{X_e^2 =
  0_{+}} & \left. {d \delta X_\parallel \over d X_e^2} \right|_{X_e^2 = 0_{-}} &= \left. {d \delta X_\parallel \over d X_e^2} \right|_{X_e^2 =
  0_{+}} \label{patchingX}\\*
  \left. \delta Z \right|_{X_e^2 = 0_{-}} &= \left. \delta Z \right|_{X_e^2 =
  0_{+}} & \left. {d \delta Z \over d X_e^2} \right|_{X_e^2 = 0_{-}} &= \left. {d \delta Z \over d X_e^2} \right|_{X_e^2 =
  0_{+}} \label{patchingZ}\,.
\end{align}
Since we expect the eigenfunction corresponding to the lowest normal
mode to be an even function of $X_e^2$, the boundary conditions that
we impose are, for the left half of the string,
\begin{align}
  \left. {d \delta X_\parallel \over d X_e^2} \right|_{X_e^2 = 0_{-}}
  = 0 &&
  \left. {d \delta Z \over d X_e^2} \right|_{X_e^2 = 0_{-}}
  = 0\label{sigmamBCs}\,.
\end{align}
The remaining challenge is to translate the boundary conditions
\eqref{sigmamBCs} into corresponding boundary conditions for $\delta
X_\parallel$ and $\delta Z$ expressed in terms of $\sigma$.

At $\sigma = \sigma_m$, the asymptotics for $\delta X_\parallel$ and
$\delta Z$ are:
\begin{align}
\delta X_\parallel(\sigma) &= a_0 + a_1 \sqrt{\sigma_m-\sigma} + a_2
(\sigma_m-\sigma) + O\left(\abs{\sigma_m-\sigma}^{3/2}\right)
\\*
\delta Z(\sigma) &= b_0 + b_1 \sqrt{\sigma_m-\sigma} + b_2
(\sigma_m-\sigma) + O\left(\abs{\sigma_m-\sigma}^{3/2}\right)\,.
\end{align}
Here,the constants $a_0$, $a_1$, $b_0$, and $b_1$ can be chosen
independently. The chain rule then gives:
\begin{align}
  {d \delta X_\parallel \over d X_e^2} = {d \sigma \over d X_e^2} {d \delta X_\parallel \over d \sigma} && {d \delta Z \over d X_e^2} = {d \sigma \over d X_e^2} {d \delta Z \over d \sigma}\,, \label{chainrule}
\end{align}
From \eqref{x2Equilibrium} we get that ${d \sigma \over d X_e^2} = C
\sqrt{\sigma_m - \sigma} +
O\left(\abs{\sigma_m-\sigma}^{3/2}\right)$ for some constant $C$,
and plugging this into \eqref{chainrule} we get
\begin{align}
  {d \delta X_\parallel \over d X_e^2} &= -{1 \over 2} C a_1 +
  O\left(\sqrt{\sigma_m-\sigma}\right)\label{Xderiv}\\
  {d \delta Z \over d X_e^2} &= -{1 \over 2} C b_1 +
  O\left(\sqrt{\sigma_m-\sigma}\right)\label{Zderiv}\,,
\end{align}
We can see from \eqref{Xderiv} and \eqref{Zderiv} that the boundary
conditions \eqref{sigmamBCs} translate into requiring $a_1 = b_1 =
0$.

In addition to the boundary conditions \eqref{BdyBCs} and
\eqref{sigmamBCs}, we should also impose a normalization condition
on $\delta X_\parallel$ or $\delta Z$, because any solution of the
above eigenvalue problem can be multiplied by an arbitrary constant
and still be a solution. We choose to impose $\delta Z'(0) = 1$. For
numerical purposes it is useful to solve \eqref{evalue} by imposing
$\delta X_\parallel(0) = \delta Z(0) = 0$, $\delta X_\parallel'''(0)
= 6 \alpha_2$, and $\delta Z'(0) = 1$, and then vary the two
parameters $\alpha_2$ and $\Omega$ until we can satisfy $a_1 = b_1 =
0$.

We find that for all the configurations to the left of the maximum
in figure~\ref{fig:lvssigmam} we can never have $a_1 = b_1 = 0$.
These, then, are stable configurations, for which the lowest normal
mode has $\omega^2>0$. The configurations to the right of the
maximum in figure~\ref{fig:lvssigmam}, however, do have a normal
mode with $\omega^2<0$, or equivalently, $\Omega^2>0$.

\begin{figure}
\begin{center}
\includegraphics[width=3.1in]{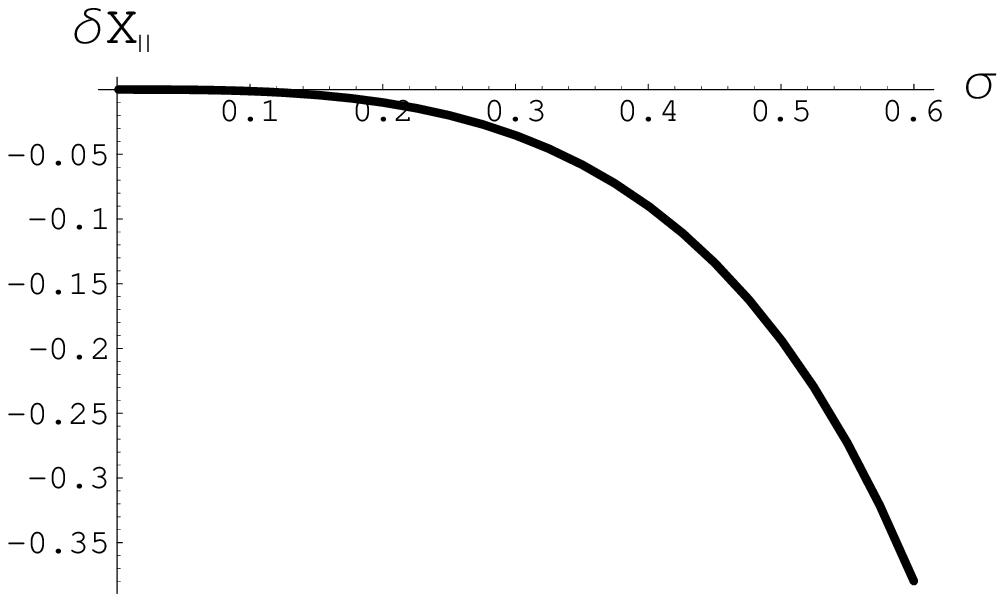}\quad
\includegraphics[width=3.1in]{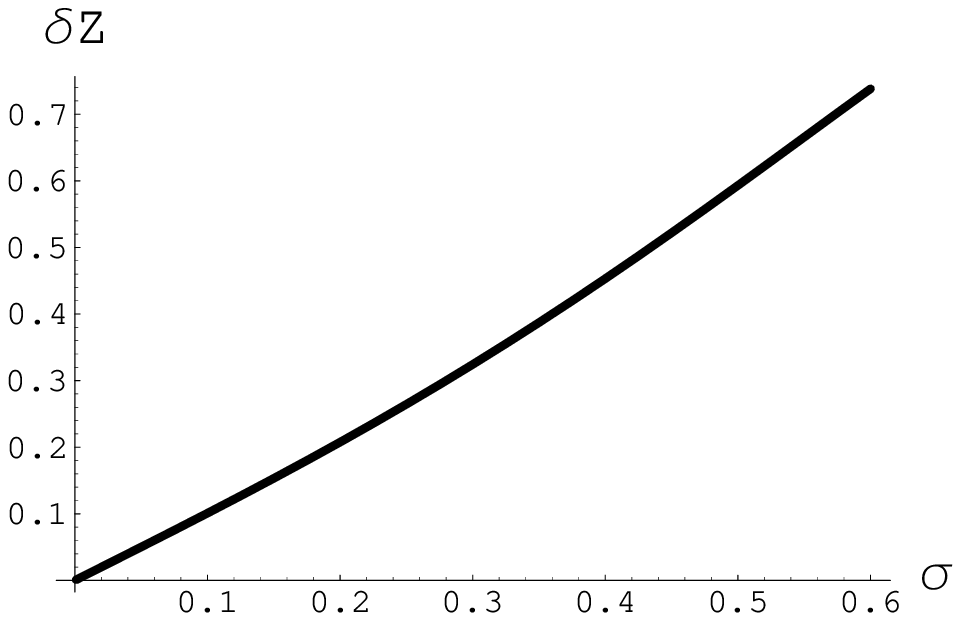}
\caption{Eigenfunctions for $v=0.9$ and $\sigma_m=0.6z_H$.  The
derivatives $\delta X_\parallel'(\sigma_m)$ and $\delta
Z'(\sigma_m)$ are finite, which means that the perturbed string
configuration is smooth at $\sigma=\sigma_m$:  see the discussion
following \eqref{sigmamBCs}.} \label{fig:efns}
\end{center}
\end{figure}

\begin{figure}
\begin{center}
\includegraphics[width=3.1in]{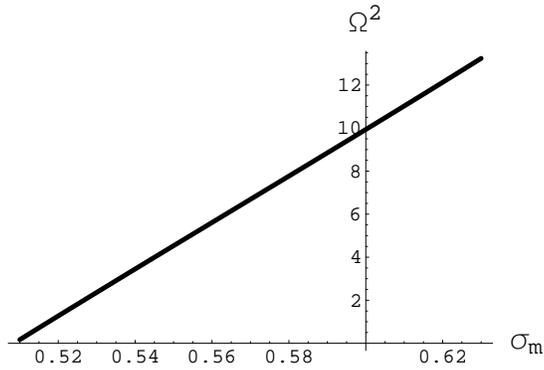}
\caption{$\Omega^2$ as a function of $\sigma_m$ for $v = 0.9$.}
\label{fig:OmegaSq}
\end{center}
\end{figure}

For example, if $v = 0.9$ the maximum $\ell_{max}$ is attained at
$\sigma_m \approx 0.51 z_H$.  If we look at a configuration with
$\sigma_m = 0.6 z_H$, we find that we can solve the eigenvalue
problem given by \eqref{evalue} with the boundary conditions
\eqref{BdyBCs} and \eqref{sigmamBCs} if we choose $\Omega \approx
3.15/z_H$.  The corresponding eigenfunctions can be seen in
figure~\ref{fig:efns}. Moreover, the dependence of $\Omega^2$ as a
function of $\sigma_m$ can be seen in figure~\ref{fig:OmegaSq}. The
dependence of $\Omega^2$ on $\sigma_m$ is evidently very close to
linear, but there are small deviations from linearity. We do not
know if these deviations are just artifacts of imprecise numerics,
or if $\Omega^2$ should actually be a linear function of $\sigma_m$.

\section{The evolution of unstable perturbations}
\label{EVOLVE}

\begin{figure}
  \centerline{\includegraphics[width=5in]{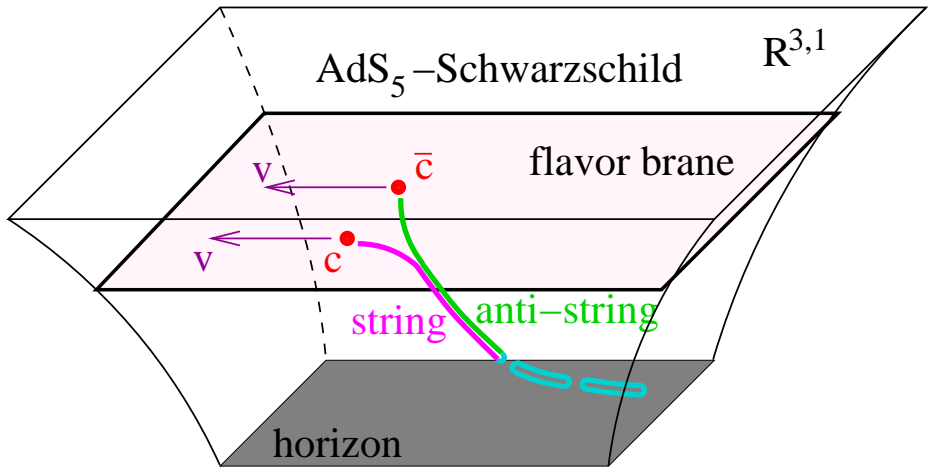}}
  \caption{A string (purple) and anti-string (green) trailing from a heavy quark and anti-quark, pictured here as $c$ and $\bar{c}$.  The string and anti-string attract, and we propose that configurations where they come together and self-annihilate are an important aspect of the dynamics.  The annihilation process of the string-anti-string pair produces highly excited closed strings, also pictured, which then fall into the horizon.  Note that the closed strings (light blue) may be of a different type from the string-anti-string pair: see the discussion in point~B below.}\label{StringPair}
 \end{figure}

For the string configurations with $\sigma_m > \sigma_{\rm max}$, which we have demonstrated to be unstable equilibria, there are several possibilities:
 \begin{enumerate}
  \item The string may evolve into the stable configuration with $\sigma_m < \sigma_{\rm max}$.
  \item The meson may dissociate, so that its constituent quarks fly away in slightly different directions, each trailing a string as in \cite{Herzog:2006gh,Gubser:2006bz}.
  \item The string may dump energy into the horizon through a mechanism involving highly excited strings.
 \end{enumerate}
The third possibility is the focus of this section.  Briefly, we propose that configurations where the string and anti-string trailing from quark and anti-quark come together into an unstable pair may play a significant role in the dynamics.  See figure~\ref{StringPair}.  The string-anti-string pair tends to self-annihilate, but depending on parameters, this may be a slow process.  If it is sufficiently slow, then based on the numerical studies of \cite{Herzog:2006gh}, one may expect that the string-anti-string pair will ``blow back'' into the trailing string configuration studied in \cite{Herzog:2006gh,Gubser:2006bz}.

In figure~\ref{StringPair} we have indicated string-anti-string annihilation by letting the trailing doubled string break into long fragments below some elevation.  This is not a process that we have good analytical control over.  But, whether it is fast or slow compared to the relaxation into the trailing string shape,
there are reasons to believe that the decay products are highly excited string states:
 \begin{enumerate}
  \item[(A)] If the coupling is weak, $g_s \ll 1$, then there is some small probability (of order $g_s^2$) per unit length of the string to break.  The result will naturally be long fragments of string, which are indeed highly excited string states.\footnote{In the limit of small $g_s$, one could hope to find a classical solution where the string and anti-string are coincident below some elevation in $AdS_5$-Schwarzschild and assume the standard trailing string shape of \cite{Herzog:2006gh,Gubser:2006bz}.  We are encouraged to learn \cite{ArgyresPrivate} that piecewise solutions have been found where the string and anti-string join at a kink.  A string-anti-string pair cannot be simply joined onto this configuration because force balance at the junction can't be maintained unless the kind turns into a cusp.  It seems likely that the string-anti-string attraction, weak though it is in the $g_s \to 0$ limit, has to play a role in the description of a configuration of the type we have depicted in figure~\ref{StringPair}.}
  \item[(B)] If the coupling is strong, $g_s \gg 1$, then we should pass to an S-dual picture where the string-anti-string pair is replaced by a D1-anti-D1 pair.  Studies of similar systems in the context of bosonic string theory \cite{Strominger:2002pc} show that in the limit where the S-dual coupling is small (meaning $g_s \gg 1$ in our original language), the unstable brane configuration decays into highly excited open strings.  Once the brane is gone, the open strings must close.  Note that in the original language, the decay products are closed D1-branes.\label{StrongCoupling}
 \end{enumerate}
In either the weak or strong coupling scenario, the highly excited strings (or D-strings) can decay into lighter states as they fall into the horizon.

Evidently, we are dealing with considerable uncertainties on the string theory side, especially if we attempt to extend the picture to intermediate values of $g_s$ (recall that $g_s = g_{YM}^2/4\pi$ is essentially $\alpha_s$, so $g_s \sim 0.3$, at least naively, for a comparison to the QGP at RHIC).  Nevertheless, it seems very likely to us that emission of massive string states plays a role in an AdS/CFT description of energy dissipation from mesons.  Another aspect of this role is that if a quark and anti-quark come close enough together, their trailing strings will attract and then annihilate into a stable meson configuration, again resulting in the emission of massive string states.

The speculative ideas discussed here could be the basis for a quantitative calculation of $\langle T_{mn} \rangle$: instead of sourcing the five-dimensional graviton with the trailing string, one should source it with pressureless dust, which is a reasonable approximation to the dynamics of massive string states.  In the case of two quarks coming together to form a meson, the initial positions and momenta of the dust particles could be specified based on the sections of the trailing string-anti-string pair they originate from.

\section{Conclusions}
\label{CONCLUSION}

We have demonstrated by an explicit analysis of linear perturbations
that one branch of Lorentzian solutions describing meson propagation
through a thermal medium is locally stable, and that the other
branch is unstable.  The stable configurations are the ones where
the string dangles less far into anti-de Sitter space:  see
figure~\ref{fig:lvssigmam}.  This stability calculation tends to
support the overall picture presented in
\cite{Peeters:2006iu,Liu:2006nn,Chernicoff:2006hi}, in which it was
argued that heavy quarkonium systems, as described in AdS/CFT, can
in certain circumstances propagate without a drag force (at least at
tree level in string theory) through the thermal medium.

We have also speculated about the possible evolution of unstable configurations of a string between a heavy quark and anti-quark.  The main substance of our remarks is that an obvious and perhaps dominant form of energy loss for such systems is the emission of massive string states.  Such states might also play a role in an AdS/CFT description of the formation of mesons out of heavy quarks.

It is essential to bear in mind that the AdS/CFT description of mesons may may or may not capture the dominant aspects of the dynamics of $J/\psi$ or $\Upsilon$ propagating through a real-world QGP.  The absence of light dynamical quarks in the AdS/CFT description is the most worrisome contrast with real-world QCD.  But even if we retreat to the most conservative position that AdS/CFT provides merely an analogous system to the QGP produced at RHIC, a more complete description of the dynamics of quark-anti-quark pairs remains an interesting avenue for future research.

\section*{Acknowledgments}

We thank P.~Argyres and M.~Edalati for stimulating discussions.  This work was supported in part by the Department of Energy under Grant No.\ DE-FG02-91ER40671, and by the Sloan Foundation.  The work of S.P.~was also supported in part by Princeton University's Round Table Fund for senior thesis research.

\clearpage

\bibliographystyle{./ssg}
\bibliography{meson}

\end{document}